\documentclass[twocolumn,a4paper,10pt]{article}
\usepackage{amsmath,amsfonts}
\usepackage{epsfig}
\usepackage{times}
\usepackage{epsfig}
\usepackage[small,hang]{caption}
\usepackage{comment}
\usepackage{graphicx}
\usepackage{subcaption}

\makeatletter  

\newcounter{numbersec}
\renewcommand{\section}[1]{\par\noindent\stepcounter{numbersec}
	\par
	\vspace{6pt}
	\noindent\textbf{\large   \arabic{numbersec} \hspace*{0.3cm} #1 }
	\par
	\vspace{2pt}
}
\renewcommand{\subsection}[1]{
	\par
	\vspace{6pt}
	\noindent\textbf{#1}
	\par
}
\renewcommand{\subsubsection}[1]{%
	\par
	\vspace{6pt}
	\textbf{#1.}
}

\newcommand{\Abstract}{\par\vspace{6pt}\noindent\textbf{\large Abstract}\par\vspace{2pt}}
\newcommand{\Acknowledgments}{\par\vspace{6pt}\noindent\textbf{\large Acknowledgments }\par\vspace{2pt}}

\newenvironment{References}{
\par\vspace{6pt}\noindent\textbf{\large References}\par\vspace{2pt}
\begin{small}\begin{list}{ }{\itemsep2mm \parsep0mm\labelsep0mm\leftmargin0mm}}
{\end{list}\end{small}}

\makeatother

\title{\vspace*{-12mm}
\LARGE \sc \textbf{  
Active flow control for three-dimensional cylinders through deep reinforcement learning
}}
\author{\vspace*{2mm}
\bf \textit{ 
P. Suárez$^{1}$, F. Alcántara-Ávila$^{1}$, A. Miró$^{2}$, J. Rabault$^{3}$, B. Font$^{2}$, O. Lehmkuhl$^{2}$ and R. Vinuesa$^{1*}$ }  \\ \\
\bf  $^{1}$ \textit{FLOW, Engineering Mechanics, KTH Royal Institute of Technology, Stockholm, Sweden} \\
\bf  $^{2}$ \textit{Barcelona Supercomputing Center - Centro Nacional de Supercomputación (BSC-CNS), Spain} \\
\bf  $^{3}$ \textit{Independent Researcher, Oslo, Norway} \\ \\
{\vspace*{3mm} \it $^*$ rvinuesa@kth.se}
}
\date{}

\begin{document}
\maketitle
\thispagestyle{empty}

%%%%%%%%%%%%%%%%%%%%
% Paper text
%%%%%%%%%%%%%%%%%%%%

%%% Insert here the abstract %%%

\Abstract

This paper presents for the first time successful results of active flow control with multiple independently controlled zero-net-mass-flux synthetic jets. The jets are placed on a three-dimensional cylinder along its span with the aim of reducing the drag coefficient. The method is based on a deep-reinforcement-learning framework that couples a computational-fluid-dynamics solver with an agent using the proximal-policy-optimization algorithm. We implement a multi-agent reinforcement-learning framework which offers numerous advantages: it exploits local invariants, makes the control adaptable to different geometries, facilitates transfer learning and cross-application of agents and results in significant training speedup. In this contribution we report significant drag reduction after applying the DRL-based control in three different configurations of the problem.

%In this contribution we report significant drag reduction under control. For $L_{\rm{jet}}/D=1$ and $Re=100$, 200, 300 and 400 we obtain $9.4\%$, $17.2\%$, $6.7\%$, $9.9\%$ drag reduction, respectively. For $L_{\rm{jet}}/D$=0.4, the results are $4.3\%, 11.1\%, 10.8\%, 15.1\%$ with local observation and $8\%, 12.7\%, 15.2\%, 11\%$ extending the observation.

%%%  Insert here the actual article text %%% 

\section{Introduction} %%%%%%%%%%%%%%%%%%

%   why? MOTIVATION FOR STUDY DRAG REDUCTION
%   Introduction of Machine learning
%   how?we are implementing ML --> DRL --> MARL needed
%   what? explain setup cylinder + previous work
%   where? ALYA + KTH CLUSTERS
%   Results are succesful 

Recent advances in the aerospace community demonstrate a growing interest in exploring new strategies for reducing emissions generated by the aviation industry. The implementation of active-flow-control systems, which aim to reduce drag, plays a vital role in the search for sustainable solutions that can effectively reduce fuel consumption, mitigate pollution and minimize vehicle transport emissions. Over the past decades, the industry has witnessed the deployment of flow-control techniques, encompassing both passive and active approaches.

One common example of passive flow control is the use of winglets on aircraft. Winglets reduce lift-induced drag on the entire wing, resulting in improved fuel efficiency and overall drag reduction. On the other hand, active control involves dynamic strategies to manipulate the flow. Synthetic jet actuators are one example of active-flow-control devices. Controlled bursts of air are potentially leading to drag reduction, improved lift-to-drag ratios and enhanced aircraft performance.

Machine-learning (ML) techniques have emerged as a valuable tool, offering the potential to uncover novel strategies highly relevant to the aerospace sector. Deep reinforcement learning (DRL) and neural networks have particularly demonstrated great promise, enabling the development of effective control strategies at a reasonable computational cost. Some recent research leveraging DRL has carried out for two-dimensional (2D) cylinders at low Reynolds numbers and known geometries actuated by zero-mass rate jets in the surface (Rabault et al. (2019), Tang et al. (2020)), and also slightly higher Reynolds numbers (Varela et al. (2023)). For a more detailed understanding of recent advances in flow control, we refer to, \textit{e.g.}, Vignon et al. (2023b) or Brunton \& Noack (2015).

DRL is based on maximizing a reward function ($R$), which is provided to an agent that interacts continuously with an environment through several action ($A$) inputs. The agent receives information about the environment state at each actuation step thanks to partial observations ($O_{st}$) of the system. Note that a sequences of consecutive actions is denoted as episode. When a batch of episodes is finished, the agent updates the neural-network weights in order to progressively determine a configuration that yields the maximum expected reward accumulated in time, for a given observation state.  

The primary objective of this study is to extend the knowledge gained from successful studies where DRL is applied to 2D cylinders (Varela et al. (2023)), turbulent channels (Guastoni et al. (2023)), and Rayleigh-Bénard convection problems (Vignon et al. (2023a)) to the scenario of three-dimensional (3D) cylinders equipped with multiple actuators on their surfaces. In this new setting, the agent observes the transition from laminar to turbulent flow in the cylinder wake and devises strategies to exploit structures of different spanwise wavelengths. As a result, each interaction of the agent is associated with a high computational cost because it needs to solve larger numerical problem per interaction. Therefore, throughout this work, striking a balance between acceptable training times and achieving optimal control performance remains a key consideration.

\section{Methodology} %%%%%%%%%%%%%%%%%%

%This section is divided into two parts: the problem description and domain setup, including the methodology for conducting numerical simulations and the implementation of the multi-agent reinforcement learning (MARL) framework. 

%CFD PROBLEM DESCRIPTION
%%  
\subsection{Problem configuration and numerical setup} %%%%%%%%%%%%%%%%%%

The problem consist of a 3D cylinder with a constant inlet velocity boundary condition $U_{\rm{in}}/U_\infty=1$. All lengths are normalized taking into account the cylinder diameter $D$. The fluid domain box has a streamwise length of $L_x=30D$. The height is $L_y=15D$ and the cylinder is located at $[7.5, 7.5]D$ in the $xy$ plane. Regarding the spanwise length, two configurations are investigated: $L_{z}=4D$ and $10D$, see Table~\ref{tab:batch_params}. We studied three different training setups denoted as W85, N85 and N255. The letter corresponds to the domain type:  W for wide ($L_{z}=10D$) and N for narrow ($L_{z}=4D$). The number denotes the $O_{st}$ size: 85 or 255 probes, without or with neighboring, explained in detail below.

\begin{figure}[h!]
	\begin{center}
	\includegraphics[width=1\linewidth]{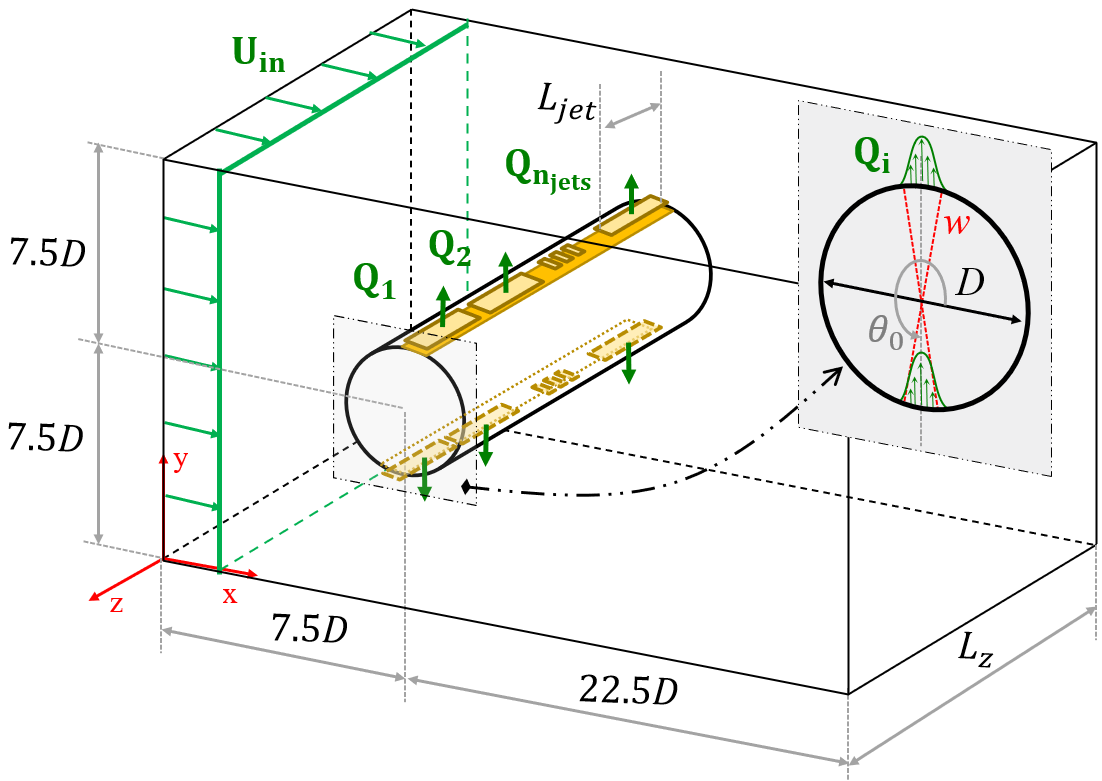}
	\caption{\label{fig:3d_domain} Non-dimensionalized configuration (reference cylinder diameter $D$). where $w$ is the jet width and $\Theta_0$ is the angular location of each jet. In green, we show the velocity condition for the inlet $U_{\rm{in}}$ and the sinusoidal profile in jets. This representation is not to scale.}
	\end{center}
\end{figure}

Regarding the rest of boundary conditions, the top, bottom and outflow surfaces (parallel to the $xz$ plane) are defined as outlets with zero velocity gradient and constant pressure. No-slip conditions are imposed in the cylinder walls, $U/U_\infty=0$, and periodic boundary conditions are imposed in the spanwise direction. The coordinate-system origin is placed in the front-face left-bottom corner, as seen in schematic representation of the domain in figure \ref{fig:3d_domain}.

The cylinder has two synthetic jets placed on the top and bottom with an arc length of $w=10^\circ$ each. These positions of the actuators are chosen to avoid the momentum injection and have a drag reduction coming from effective actuation. These jet velocities $V_{\rm{jet}}$ are a function of both the jet angle $\theta$ and the desired mass flow rate $Q$ determined by the DRL control. The jet velocity profile is defined as follows:
\begin{equation} \label{eq:jet_eq}
	V_{\rm{jet}}(Q,\omega)=Q\frac{2\pi}{\omega D^2}\cos \left [ \frac{\pi}{\omega}(\Theta-\Theta_0) \right ],
\end{equation}
where $\Theta_0$ corresponds to the angle where the jet is centered (in this problem, 90º and 270º degrees). The scaling factor is used so the integration of the jet velocity corresponds to the mass flow rate and the cosine function ensures zero velocity at the boundaries with the cylinder.

The flow profile within an individual actuator jet is constant in all its spanwise length. No spatial smoothing is needed in the arc boundary that exist between adjacent actuators.

The Reynolds number, $Re=\rho U_{in} D/\mu$, where $\rho$ is the density and $\mu$ is the molecular viscosity, considered are 100, 200, 300 and 400. This range contains the transition from laminar flow to the emergence of three-dimensional instabilities in the cylinder wake (Williamson (1996) and Zhang et al. (1995)). The motivation of this work is to assess how the control is capable of tackling and exploiting the different wake structures in 3D. 

The numerical simulations are carried out by means of the numerical solver Alya (Vázquez et al. (2016)). The spatial discretization is based on the finite-element method (FEM) and the incompressible Navier-Stokes equations are considered.

It is worth noting that, due to the large amount of training required for the DRL control considered here, the computational cost of numerical simulations dominates in determining the overall wall-clock time. When designing the mesh, a compromise between cost and accuracy has been made, ensuring that the chosen mesh effectively captures the primary structures and wavelengths in the cylinder wake. This provides the agent essential information for controlling them. 

%Once the training is completed, the strategies can be tested on higher-fidelity meshes to validate their performance.

\begin{table}[]
\fontsize{8}{12}\selectfont
\begin{tabular}{c|cccc|}
\cline{2-5}
Case                    & \multicolumn{1}{c|}{2D}       & \multicolumn{1}{c|}{W85}  & \multicolumn{1}{c|}{N85} & N255 \\ \cline{2-5} 
Mesh cells      & \multicolumn{1}{c|}{43600}                   & \multicolumn{1}{c|}{6.2M} & \multicolumn{2}{c|}{2.6M}      \\ \cline{2-5} 
$L_x/D$                     & \multicolumn{4}{c|}{30}                                    \\ \cline{2-5} 
$L_y/D$                     & \multicolumn{4}{c|}{15}                                     \\ \cline{2-5} 
$L_z/D$                     & \multicolumn{1}{c|}{-}                  & \multicolumn{1}{c|}{10}   & \multicolumn{2}{c|}{4}         \\ \cline{2-5}
$L_{\rm{jet}}/D$                     & \multicolumn{1}{c|}{-}                  & \multicolumn{1}{c|}{1}   & \multicolumn{2}{c|}{0.4}         \\ \cline{2-5}
$O_{\rm{st}}$        & \multicolumn{3}{c|}{85}                             & 255 \\ \cline{2-5} 
$Q_{\rm{max}}$                    & \multicolumn{1}{c|}{0.088}              & \multicolumn{3}{c|}{0.176}                                 \\ \cline{2-5} 
$S_{\rm{n}}$                  & \multicolumn{4}{c|}{2}                                                                               \\ \cline{2-5} 
$R_{\rm{n}}$                  & \multicolumn{4}{c|}{5}                                                                               \\ \cline{2-5} 
$T_a$                       & \multicolumn{4}{c|}{0.25}                                                                            \\ \cline{2-5} 
Actions/episodes      & \multicolumn{1}{c|}{100}                & \multicolumn{3}{c|}{120}                                   \\ \cline{2-5} 
CPUs/environment     & \multicolumn{1}{c|}{1}                  & \multicolumn{3}{c|}{128}                                   \\ \cline{2-5} 
Parallel environments & \multicolumn{1}{c|}{10}                 & \multicolumn{3}{c|}{8}                                     \\ \cline{2-5} 
Baseline duration [TU]   & \multicolumn{4}{c|}{150}                                                                             \\ \cline{2-5} 
Lift penalty             & \multicolumn{4}{c|}{0.6}                                                                             \\ \cline{2-5} 
\#Neurons(layers)              & \multicolumn{4}{c|}{512(2)}                                                                         \\ \cline{2-5} 
Reynolds numbers         & \multicolumn{4}{c|}{100, 200, 300 and 400}                                                           \\ \cline{2-5} 
Time smoothing  & \multicolumn{1}{c|}{linear}             & \multicolumn{3}{c|}{exponential}                           \\ \cline{2-5} 
\end{tabular}
\caption{\label{tab:batch_params} Main parameters of the simulations for each training setup and compared with the benchmark in 2D.}
\end{table}

%DRL FRAMEWORK
%%
\subsection{Multi-agent reinforcement learning (MARL)} %%%%%%%%%%%%%%%%%%

Previous work done in 2D cylinders used single-agent reinforcement learning (SARL), where every set of actions is decided at once. Note that with the increase of the action space, this becomes a much more challenging task because it is necessary to find the best policy for a high-dimensional control. Therefore, SARL is not viable option in 3D cylinders because the agent need more episodes to tackle all possible combinations in the $n$ jets located in the cylinder surface. Added to the fact that the computational cost per action is orders of magnitude higher than 2D environments then the total wall-clock time required becomes excessive. On the other hand, the potential of MARL in these cases has been recently documented by Belus et al. (2019) and Vignon et al. (2023a).

The MARL framework avoids the curse of dimensionality present in this particular setup. This new approach, in contrast to SARL, aims to train locally on environment partitions, which are denoted as pseudo environments. Note that all of them share the same neural-network weights. Doing so, the high-dimensional control space becomes tractable and the agent is trained in smaller domains to maximize the local rewards, some additional features are added to ensure the pursue of global reward maxima. 

\begin{comment}
\begin{figure}[h!]
	\begin{center}
	\includegraphics[width=1\linewidth]{drl_schemes/MARL_sketch.png}
	\caption{\label{fig:MARL_SKETCH} Multi-agent reinforcement learning (MARL) coupled with a multiple-actuator 3D cylinder environment. Three main communications are represented, $O_{st} = S_i$, $R=R_i$ and $a=A_i$}
	\end{center}
\end{figure}
\end{comment}

The agent interacts with the numerical simulation domain through three main channels. The observation state $O_{st}$, that is sent from the simulation to the agent, consists of partial pressure information from slices of 85 probes in the wake, centered on the corresponding pseudo environment location in $z$ (not shown here). In the present work, two configurations are considered, \textit{i.e.} with or without observation of neighboring pressure values, as shown in table \ref{tab:batch_params}. The neighboring consist in adding slices of each side, one set of 85 probes per side. The observation state becomes three slices of probes, 255 pressure values in total. 

The simulation also sends the DRL environment total reward $R$, see Equation (\ref{eq:reward_eq}) below, defined as a sum of the local and global reward. The scalar $R_{\rm{n}}$ fit the reward signal into $[-1,1]$ range as Tensorforce libraries require. A new heuristic parameter is added, the parameter $\beta$, used to balance the local and global rewards, in this research is set to $\beta=0.8$.  

These rewards $r$, Equation (\ref{eq:reward2_eq}), are computed as a function of the drag coefficient reduction, $\Delta C_d = C_d(t,i_{\rm{jet}})-C_{d_{\rm{b}}}$, being $C_{d_{\rm{b}}}$ the uncontrolled baseline known value. In addition we have a lift contribution multiplied by $\alpha$, acting as a penalty to avoid axis-switching and ensure only reduction in the streamwise force component.  

Note that aerodynamic forces ($C_d$ and $C_l$) are defined in Equation (\ref{eq:forces_eq}). The frontal area $A_f$ corresponds to local pseudo environment surfaces for $C_{{d,l}_{\rm{local}}}$ and the whole cylinder surface for $C_{{d,l}_{\rm{global}}}$. 

% Increase the vertical space between lines in multline
\setlength{\multlinegap}{4em} % Adjust the value as needed

\begin{equation} \label{eq:reward_eq}
    \fontsize{9}{3}\selectfont
	R(t,i_{\rm{jet}})=R_{\rm{n}}(\beta r_{\rm{local}}(t,i_{\rm{jet}}) + (1-\beta)r_{\rm{global}}(t)),
\end{equation}
\begin{equation} \label{eq:reward2_eq}
    \fontsize{9}{3}\selectfont
    r(t,i_{\rm{jet}})=C_{d_{\rm{b}}}-C_d(t,i_{\rm{jet}})-\alpha\vert C_l(t,i_{\rm{jet}})\vert,
\end{equation}
\begin{equation} \label{eq:forces_eq}
    \fontsize{9}{3}\selectfont
    \text{where} \quad C_d=\frac{2 F_x}{\rho A_f v_{\infty}^{2}}  \quad \text{and} \quad C_l=\frac{2 F_y}{\rho A_f v_{\infty}^{2}}.
\end{equation}

The action $A$ is computed by the agent based on the state of the system. The DRL library employed here outputs this value in the range $[-1,1]$, thus this value needs to be rescaled as $Q=A Q_{\rm{max}}$ in order to avoid excessively large actuations. During training we observed that the $Q_{\rm{max}}$ obtained in 2D studies were not adequate in the context of the present 3D cylinders. Thus, $Q_{\rm{max_{3D}}}=2Q_{\rm{max_{2D}}}=0.176$ was set to yield adequate results. Note that, based on Equation (\ref{eq:jet_eq}), $Q$ is directly related to the mass flow rate from the jet. For each pseudo environment, we set is opposite action values between the top and bottom jet, \textit{i.e.} $Q_{\rm{90^\circ}}=-Q_{\rm{270^\circ}}$, in order to ensure the global zero mass flow rate. Although one needs to take into account the energy consumption of the actuator in order to calculate the net energy saving, this is highly dependent on the actual experimental setup. In the present numerical setting the cost of the control is negligible compared with the drag reduction (Guastoni et al. (2023)), this would not necessarily be true in an experiment.

Every action $A$ from the agent is applied in the system during $T_a$ time units. Jet boundary conditions are updated following Equation (\ref{eq:jet_eq}). The transition in time between actions, $Q_{t} \rightarrow Q_{t+1}$, is done by an exponential function in time.  

Some DRL setup parameters are closely related to the fluid-mechanics problem at hand. The duration of an episode is defined to contain 6 vortex-shedding periods ($T_k=1/f_k$). In this case, the Strouhal number for the range of Reynolds numbers under consideration is around $St=f_kDU_{\rm{in}}/U_\infty=0.2$. Note that we set $T_a<0.05T_k$, which is in agreement with the recommendations from previous publications. Consequently, a total of 120 actuations per episode is considered to be adequate to evaluate the accumulated reward. Also note that every episode starts from a uncontrolled converged state of the problem.  

The neural-network architecture consists of two dense layers of 512 neurons. A proximal-policy-optimization agent define the neural-network weights based on policy-gradient method. The open-source library Tensorforce is used (Kuhnle et al. (2017)). The batch size, \textit{i.e.} the total experiences uses the PPO agent for each gradient-descent iteration, is set to 80. This is different from the standard size of 20 used in previous implementations. This has been modified for computational cost and multi-environment synchronization purposes, being an adequate configuration to run enough experiences at the same time to do neural-networks updates efficiently. If 8 environments (independent simulations) with 10 pseudoenironments each are running at the same time, it is essential to not lose any information when the next $10\times 8$ experiences begin. Consequently, the next episodes will not start until the neural-network weights are updated. 

It is important to mention that there is an individual agent for each Reynolds number and case setup. Although transfer-learning techniques have shown good potential, they are not applied in the present work because the focus here is to compare setups with MARL and define benchmark focused on assessing how the agents discover approaches to control wake instabilities.

%% not talking about COMPUTATIONAL COST?
%% multienvironment setup? 
%% batch size explanation? 

% previous concept of SARL -- and direct application not feasiblei
% new variant MARL to exploit local invariants and be "cheaper"/feasible, cite COLIN REVIEW
% cite previous succesful work by BELUS 2019 and COLIN 2023
% comment how good suits the 3D cylinder problem --> extrapolate, physics, flexible in geometries, speedup in trainings proportional to the pseudo environments number per CFD enviroment
% difference table --> define witnesses, problem with neighboring and length jet
% changes from varela --> smoothing, baseline duration, influence of the Qmax, reward weight in the local and global drag, 

\section{Results and discussion} %%%%%%%%%%%%%%%%%%

To the authors' knowledge, this study constitutes the first time that a 3D cylinder with multiple jets configuration is successfully trained using MARL.

\begin{figure}[h!]
	\begin{center}
	\includegraphics[width=1\linewidth]{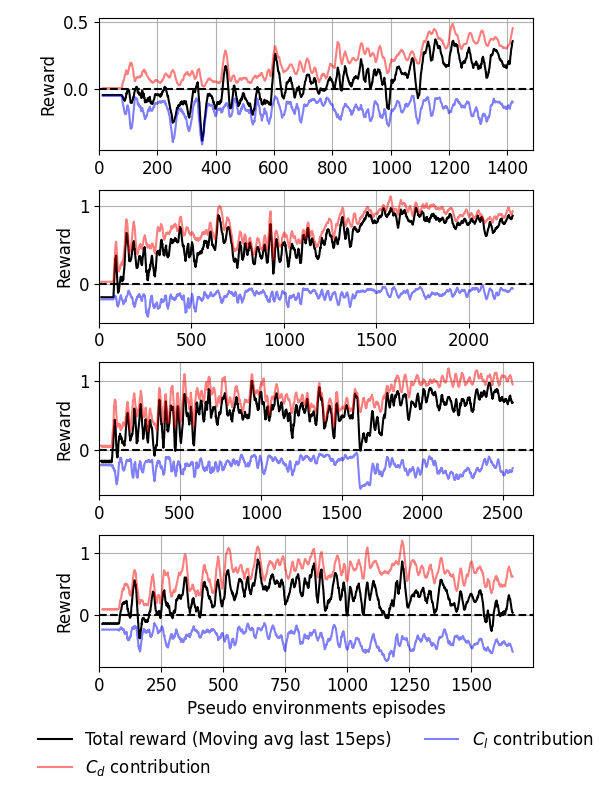}
	\caption{\label{fig:train_batch1} Training curves showing the reward in N255 case for $Re=100$, $200$, $300$ and $400$ (from top to bottom)}
	\end{center}
\end{figure}

Figure \ref{fig:train_batch1} shows all pseudo environment rewards $R$, together with the pure drag reduction and lift-biased penalisation. For instance, in the $Re=200$ case, the lift contribution to the reward is close to zero for the later episodes, a fact that indicates that the agent has discovered a very good control approach. Also note that, even if the reward remains close to 0, the agent may be learning. This can be observed in the $Re=400$ case, where a strategy with great drag reduction is achieved, although at the cost of inducing lift biases. The aerodynamic impact of such an effect needs to be accounted for when assessing the merit of the control approach.

\begin{figure}[h!]
	\begin{center}
	\includegraphics[width=1\linewidth]{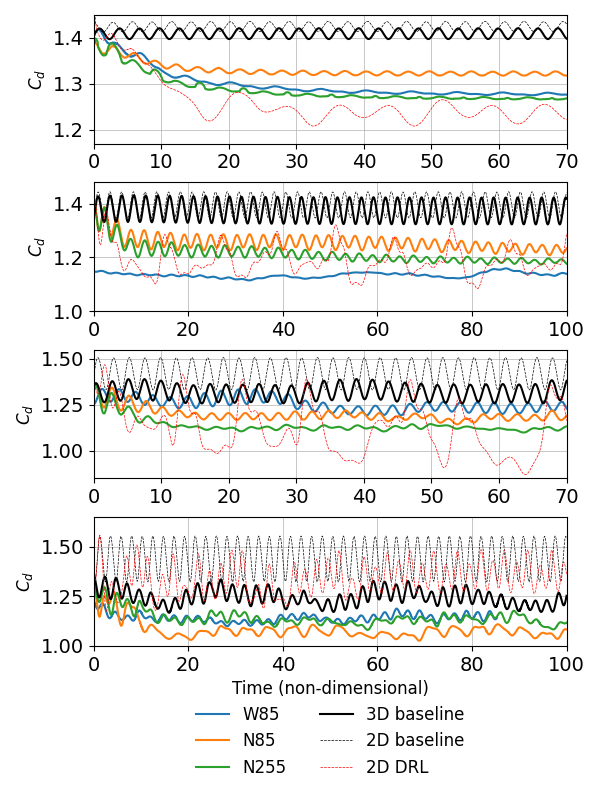}
	\caption{\label{fig:CD_ALL} Evolution of the drag coefficients as a function of time for all training cases (from top to bottom): $Re=100$, $200$, $300$ and $400$.}
	\end{center}
\end{figure}

After finishing the various training cases, the performance of the agent is evaluated by running the obtained policy in deterministic mode. In this case, the policy is evaluated without any exploration. Consequently, the agent computes the most probable value of the action $A$ probability distribution that ensures the maximum expected reward. The case runs until the control converges into a periodic control behaviour. All the cases lead to effective drag reduction rates. The drag-reduction rates reported for all the 3D cases in Table \ref{tab:cd_reduction} are slightly different from those obtained 2D. In the latter case the physics is significantly constrained, and as expected the discrepancy between 2D and 3D results increases with $Re$ (both in controlled and uncontrolled cases). When comparing the drag coefficient signals (Figure \ref{fig:CD_ALL}), we observe that the performance of the 3D control strategies are more consistent for increasing $Re$ than that of the 2D cases. Thus, while the performance of the 2D model at $Re=400$ is degraded compared with the low-$Re$ case, the 3D case still exhibits excellent performance. Note that all results are presented in dimensionless units.

\begin{table}[h!]
\centering
\fontsize{9}{11}\selectfont
\begin{tabular}{ccccc}
\hline
$Re$                   & 2D benchmark & W85 & N85 & N255 \\ \hline
100                  & 13.0       & 9.4     & 4.3     & 8.0        \\
200                  & 14.9       & 17.2    & 11.1    & 12.7         \\
300                  & 21.9       & 6.7     & 10.8    & 15.3         \\
400                  & 5.6        & 9.9     & 15.1    & 11.1        
\end{tabular}
\caption{\label{tab:cd_reduction} Summary of of percentual drag reduction $\left[ (1 - C_{d_{\rm{DRL}}}/{C_{d_{\rm{b}}}}) \times 100\% \right]$ obtained in deterministic converged stages for each case.}
\end{table}

\begin{figure}[h!]
    \subfloat[$Re=300$ from N85 case]{\includegraphics[width=1\linewidth]{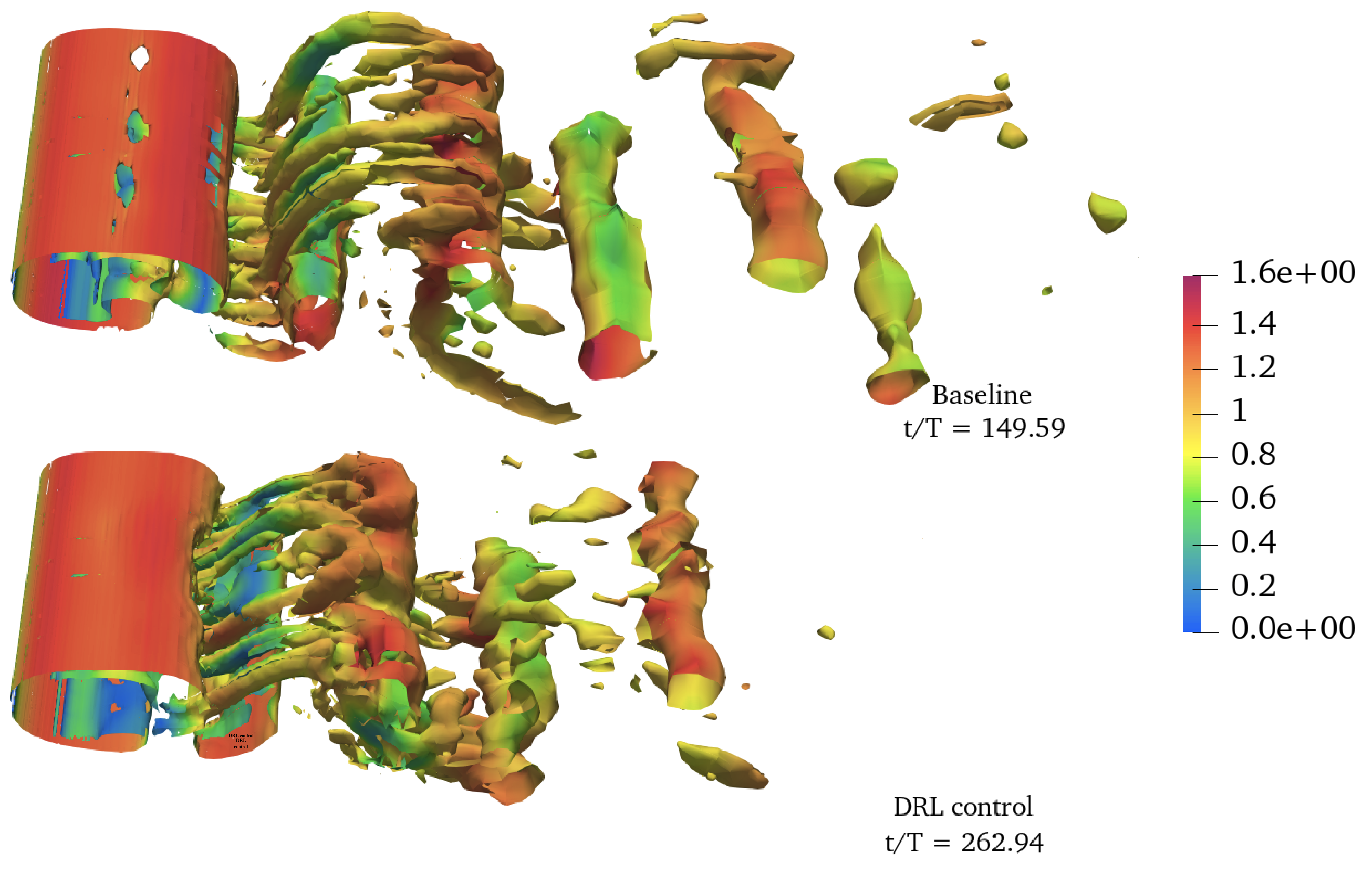}}
    \hfill
    \subfloat[$Re=400$ from N85 case]{\includegraphics[width=1\linewidth]{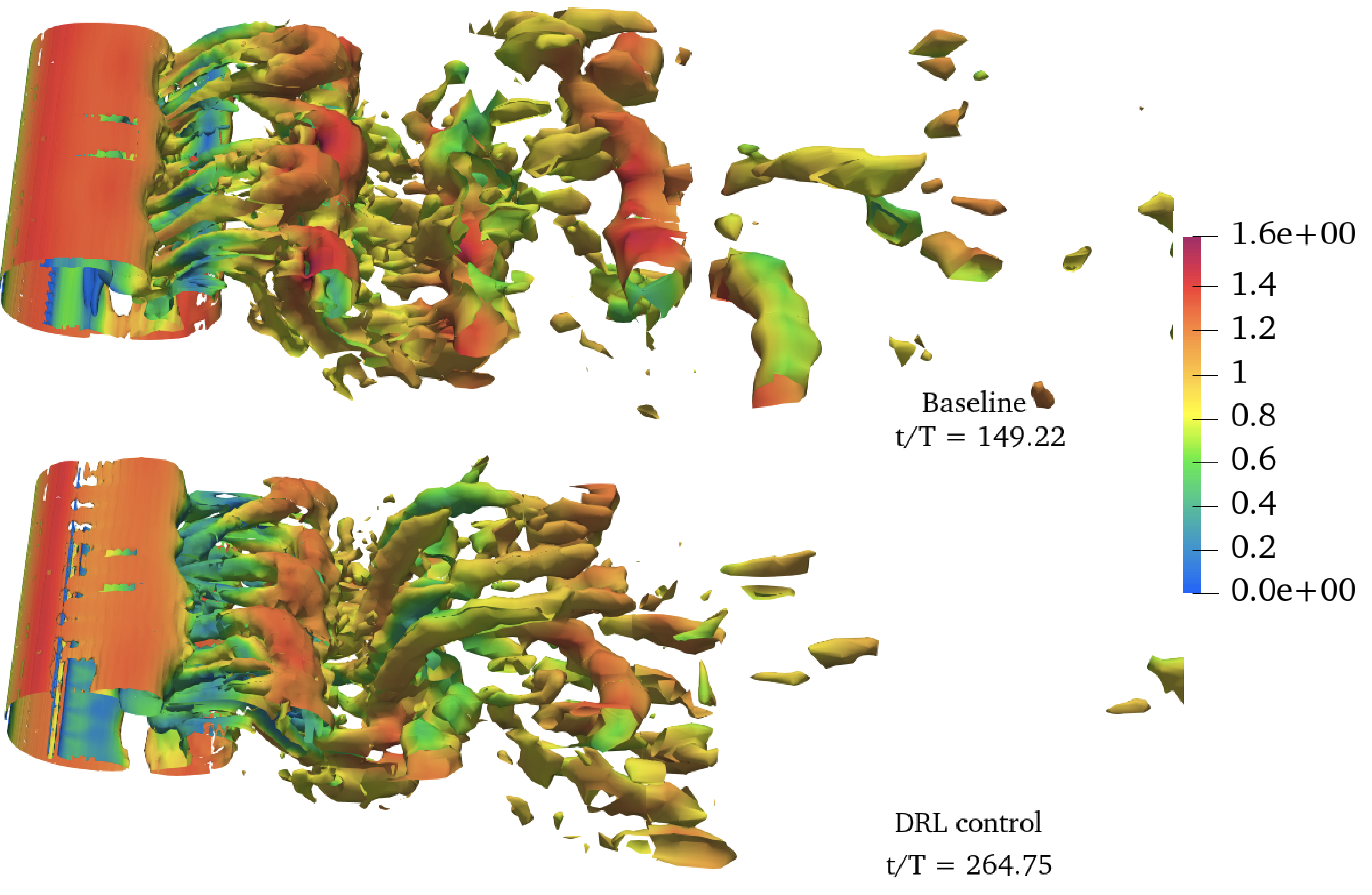}}
    \caption{Instantaneous snapshots comparing the baseline (top) and controlled (bottom) cases. We show vortical motions (Hunt (1987)) defined by isosurfaces equal to (a) 0.5 and (b) 0.35, colored by streamwise velocity.}
    \label{fig:q_crit}
\end{figure}

As shown in Figure \ref{fig:q_crit}, the DRL control leads to an attenuation of the vortex-shedding strength, as illustrated by visualizing vortical structures (Hunt (1987)). Also, note that the control give rise to vortex-street instabilities earlier in $Re=200$ than expected in the uncontrolled cases (not shown here).

\begin{figure}[h!]
	\begin{center}
	\includegraphics[width=1\linewidth]{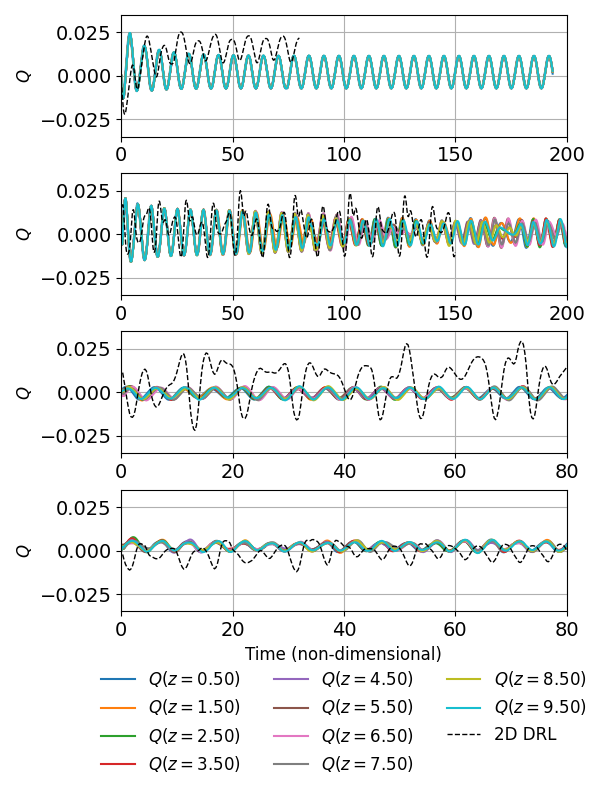}
	\caption{\label{fig:MFR1} Actions Q given by agent corresponding to each jet (Equation (\ref{eq:jet_eq})) for the following cases in W85 (from top to bottom): $Re=100$, $200$, $300$ and $400$.}
	\end{center}
\end{figure}

The time series of the actions $Q$ for the various cases are shown in Figure \ref{fig:MFR1}. We note that, for $Re=200$, between $t=150$ and 200 there is a noticeable change in the amplitude of the signal. This shows how the policy is able to exploit combinations of $Q$ while avoiding the curse of dimensionality. In an overall analysis the blowing intensity is different in 3D compared to 2D because the physical system is different as we increase the Reynolds number. 
%This is quite expected because the constrained 2D domain has more sensitivity to control.

However, the actual control in most of the 3D cases presented seem to have an "extruded" strategy. All jets blow in sync and can be simplified as a constant velocity profile along the cylinder span. Maybe the three-dimensional instabilities appearing are weak enough to not dominate in the near wake. The actions may not take advantage yet because the low $Re$ regime or the configuration studied here. Our results suggest that shorter jets in spanwise direction may be better to deal with higher Reynolds numbers, although this point will be further assessed in future work. Maybe the cylinders studied are short in order to see bigger patterns in the control. Smaller $L_{\rm{jet}}$ and better $O_{\rm{st}}$ can be key to discover non-"extruded" control that may yield to the wake instabilities exploitation. Note that the narrower jets receive a more local observation and can in principle exploit position of the wake structures (modes A and B as seen in literature Williamson (1996)) to develop policies leading to higher drag reduction. In future work we will investigate if this "extruded" strategy is best in general or if a more sophisticated control can be observed in different setup configurations.

\begin{comment}
JEAN --------
- the blowing intensity / phase is different in 3D compared to 2D (because the sensitivity to control / "receptivity" of the physical system is different in 2D and 3D): this is quite expected / intuitive
- however, the actual control strategy in 3D is mostly an "extruded" strategy; it seems that all the jets more or less blow in sync. This is maybe a bit more surprising at first. This means that the control strategy is mostly 3D. You can speculate on this: maybe this is because the Re is low enough that the 3D structures are weak enough to not be the dominating feature in the near wake? or maybe the cylinder is still too "short" to allow patterns to emerge in which smart control can take stronger advantage of the 3-dimensionality of the flow? This will be tested in the future by looking into i) longer cylinders, ii) higher Re, and we will investigate further if such an "extruded" strategy is best in general, or if a more sophisticated strategy can be observed in different conditions.
\end{comment}

\begin{figure}[h!]
	\begin{center}
	\includegraphics[width=1\linewidth]{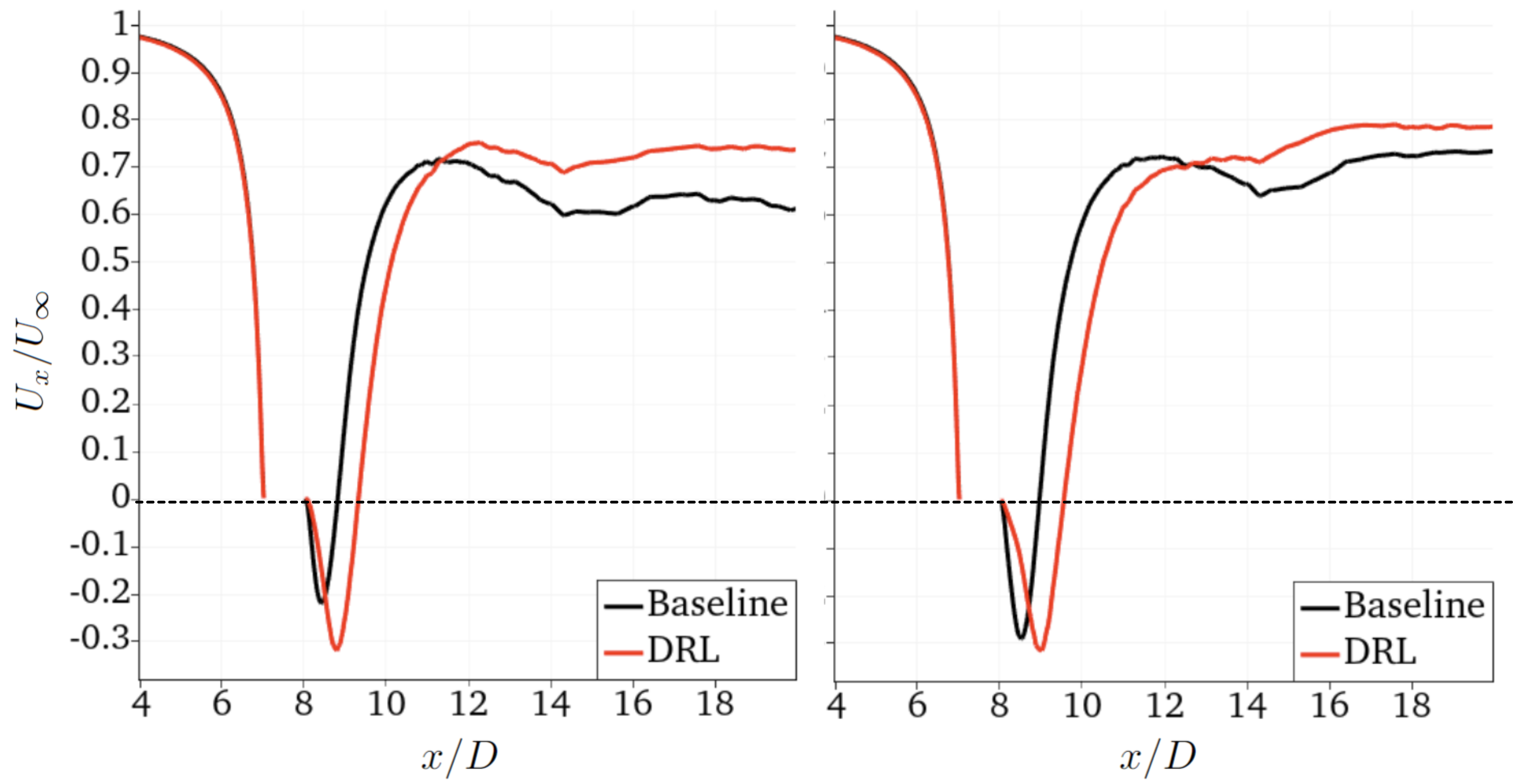}
	\caption{\label{fig:avg_bubble} Mean streamwise velocity at y = 7.5D (cylinder center). We show N85 case at (left) $Re=300$ and (right) $Re=400$. The dashed line indicates $U_x/U_\infty=0$.}
	\end{center}
\end{figure}

The recirculation bubble downstream of the cylinder is studied through the mean streamwise velocity in Figure \ref{fig:avg_bubble}. This figure indicates that the reattachment location is delayed in the controlled cases, which exhibit a higher velocity than the uncontrolled one for larger $x/D$, a fact that indicates that the wake is less affected by the bluff body.

\begin{comment}
\begin{figure}[h!]
	\begin{center}
	\includegraphics[width=1\linewidth]{coeffs_deterministic/mfr_ETMM_batch1.png}
	\caption{\label{fig:MFR2} Actions Q given by agent corresponding to mass flow rate (Equation \ref{eq:jet_eq}) for $L_z$=4D and $L_{\rm{jet}}$=0.4D (Batch \#2)}
	\end{center}
\end{figure}

\begin{figure}[h!]
	\begin{center}
	\includegraphics[width=1\linewidth]{coeffs_deterministic/mfr_ETMM_batch2.png}
	\caption{\label{fig:MFR2} Actions Q given by agent corresponding to mass flow rate (Equation \ref{eq:jet_eq}) for $L_z$=4D and $L_{\rm{jet}}$=0.4D (Batch \#2)}
	\end{center}
\end{figure}
\end{comment}

\begin{comment}
\begin{figure}[h!]
	\begin{center}
	\includegraphics[width=1\linewidth]{qcrit/qcrit_re100_22.pdf}
	\caption{\label{fig:q_crit22} Instantaneous between baseline (top) and controlled(bottom) snapshop isosurface at $Q-criterion$=0.35 corresponding at $Re$=100 training batch \#3}
	\end{center}
\end{figure}

\begin{figure}[h!]
	\begin{center}
	\includegraphics[width=1\linewidth]{qcrit/qcrit_re200_19.pdf}
	\caption{\label{fig:q_crit19} Instantaneous between baseline (top) and controlled(bottom) snapshop isosurface at $Q-criterion$=0.5 corresponding at $Re$=200 training batch \#2}
	\end{center}
\end{figure}
\end{comment}

\section{Conclusions} %%%%%%%%%%%%%%%%%%

In this study, the MARL framework is coupled with numerical solver Alya to train and find optimal drag-reduction strategies, controlling multiple jets placed in the spanwise direction of a 3D cylinder. Recent state-of-the-art studies in DRL control in 2D cylinders has been extended with new implementations to account for the wake three-dimensionality. This study is carried out in the transition regime where vortex-street instabilities emerge, and this constitutes an additional challenge for DRL at various Reynolds numbers. Our results indicate that MARL is essential to achieve learning in the cases under study by exploiting the underlying physics within pseudo environments and optimizing the global problem involving multiple interactions in parallel. Further investigations will be carried out into the exploration of the action space size and jet dimensions. One of the main advantages of using MARL is the ability to deploy the trained agents for different cylinders lengths and actuator numbers just maintaining $L_{\rm{jet}}$. Note that the training focuses on the symmetries and invariant structure through all spanwise direction. This would not be possible with SARL, which is restricted to a certain number of actuators. Furthermore, MARL allows performing cheaper training sessions in smaller and under-resolved domains, speeding up the process, before tackling the control in high-fidelity simulations. 

The training results demonstrate effective control for $Re$=100, 200, 300, and 400, achieving drag reductions of 9.4\%, 17.2\%, 6.7\%, and 9.9\%, respectively, when using a jet length of $L_{\rm{jet}}/D$=1. For a jet length of $L_{\rm{jet}}/D$=0.4, the drag reduction is 4.3\%, 11.0\%, 10.8\%, and 15.08\% with local observation, and 8.0\%, 12.7\%, 15.2\%, and 11\% when extending the observation to spanwise neighbors. These findings highlight the effectiveness of the training process in achieving significant drag reduction across different cases with slightly different DRL configurations. 

Future work will leverage the present coupling between MARL and AFC problems for more realistic cases, scaling up to turbulent regimes and more complex geometries. 

 % comment about % cd reduction
 % most important conclusions about the control in 3D 
 % MARL is all we need
 % Future work --> study statistics, physics and parametric study of L_jet
 % Very powerful applications 

Furthermore, the present results bring new benchmark results for the DRL community, which may motivate its use for future applications.
%%% Insert here acknowledgments if necessary %%%

\Acknowledgments

Ricardo Vinuesa acknowledges funding by the ERC through Grant No. ``2021-CoG-101043998, DEEPCONTROL''.

%%% References %%%

% BELUS
% COLIN RBC
% COLIN REVIEW
% WILIIAMSON 1996
% ZHANG 1995
% VARELA
% XIAN-JUN

\begin{References}

\item Belus, V., Rabault, J., Viquerat, J., Che, Z., \& Hachem, E. (2019), Exploiting locality and translational invariance to design effective deep reinforcement learning control of the 1-dimensional unstable falling liquid film. AIP Advances, 9, 125014.

\item Brunton, S. L., \& Noack, B. R. (2015). Closed-Loop Turbulence Control: Progress and Challenges. ASME. Appl. Mech. Rev. September 2015; 67(5): 050801.

\item Guastoni, L., Rabault, J., Schlatter, P., Azizpour, H., \& Vinuesa, R. (2023), Deep reinforcement learning for turbulent drag reduction in channel flows. Eur. Phys. J. E 46, 27.

\item Hunt, J.C.R (2018), Vorticity and Vortex Dynamics in Complex Turbulent Flows. Transactions of the Canadian Society for Mechanical Engineering. 11(1): 21-35.

\item Kuhnle, A., Schaarschmidt, M., \& Fricke, K. (2017), Tensorforce: A TensorFlow library for applied reinforcement learning. Web page, 9.

\item Rabault, J., Kuchta, M., Jensen, A., Réglade, U., \& Cerardi, N. (2019), Artificial neural networks trained through deep reinforcement learning discover control strategies for active flow control. Journal of Fluid Mechanics, 865, 281-302.

\item Tang, H., Rabault, J., Kuhnle, A., Wang, Y., \& Wang, T. (2020), Robust active flow control over a range of Reynolds numbers using an artificial neural network trained through deep reinforcement learning. Physics of Fluids, 32(5), 053605.

\item Varela, P., Suárez, P., Alcántara-Ávila, F., Miró, A., Rabault, J., Font, B., García-Cuevas, L. M., Lehmkuhl, O., \& Vinuesa, R. (2022), Deep reinforcement learning for flow control exploits different physics for increasing Reynolds number regimes. Actuators, 11(12), 359.

\item Vázquez, M., Houzeaux, G., Koric, S., Artigues, A., Aguado-Sierra, J., Arís, R., Mira, D., Vázquez, H., Cucchietti, F., Owen, H., Taha, A., Burness, E., Cela, J., \& Valero, M. (2016), Alya: Multiphysics engineering simulation toward exascale. Journal of computational science, 14, 15-27.

\item Vignon, C., Rabault, J., Vasanth, J., Alcántara-Ávila, F., Mortensen, M., \& Vinuesa, R. (2023a), Effective control of two-dimensional Rayleigh--Bénard convection: Invariant multi-agent reinforcement learning is all you need. Physics of Fluids, 35, 065146

\item Vignon, C., Rabault, J., \& Vinuesa, R. (2023b), Recent advances in applying deep reinforcement learning for flow control: Perspectives and future directions. Physics of Fluids, 35(3), 031301.

\item Williamson, C. H. K. (1996), Vortex dynamics in the cylinder wake. Annual Review of Fluid Mechanics, 28, 477-539.

\item Zhang, H., Fey, U. B. R., Noack, K., König, M., \& Eckelmann, H. (1995), On the transition of the cylinder wake. Physics of Fluids, 7(4), 779-794.

\end{References}

\end{document}